\documentclass[twocolumn,showpacs,showkeys,preprintnumbers,amsmath,amssymb]{revtex4}

\usepackage{graphicx}
\usepackage{dcolumn}
\usepackage{bm}

\begin{document}

\title{Orbital Ordering and Spin-Ladder Formation in
       $ {\rm \bf La_2RuO_5} $}

\author{V.\ Eyert}
\author{S.\ G.\ Ebbinghaus}
\author{T.\ Kopp}

\affiliation{Institut f\"ur Physik, Universit\"at Augsburg,
             86135 Augsburg, Germany}

\date{\today}

\begin{abstract}
The semiconductor-semiconductor transition of $ {\rm La_2RuO_5} $
is studied by means of augmented spherical wave (ASW) electronic
structure calculations as based on density functional theory and
the local density approximation. This transition has lately
been reported to lead to orbital ordering and a quenching of the
local spin magnetic moment. Our results
hint towards an orbital ordering scenario which, markedly different
from the previously proposed scheme, preserves the local $ S = 1 $
moment at the Ru sites in the low-temperature phase. The
unusual magnetic behaviour is interpreted by the formation of
spin-ladders, which result from the structural changes occurring
at the transition and are characterized by antiferromagnetic
coupling along the rungs.
\end{abstract}

     % Exchange interactions

\maketitle

%\section{Introduction}

The orbital degeneracy of $ d $-shell atoms lays ground for
numerous exciting phenomena observed in transition-metal compounds
\cite{imada98}. Orbital fluctuations as well as orbital ordering
lead to extraordinary ground states, low-energy
excitations and phase transitions. Well known examples for such
ordering phenomena are the perovskite-based manganites
\cite{rao98,manganates},
and the antiferro-orbital structure in $ {\rm KCuF_3} $
\cite{liechtenstein95}.
Increased complexity is observed for compounds, where the orbitals
couple to the spin or charge degrees of freedom. This situation has
been studied by Kugel and Khomskii for magnetic systems
\cite{kugel}. Orbital and magnetic ordering has also been found in
the triangular chain magnet $ {\rm Ca_3Co_2O_6} $, which is
characterized by an alternation of
low-spin and high-spin sites \cite{cobaltates}. In contrast,
interplay of charge and orbital order as well as singlet formation
has been demonstrated to play a significant role in the Magn\'{e}li
phase $ {\rm Ti_4O_7} $ \cite{titanates}.
While orbital ordering has been primariliy studied for transition-metal 
compounds of the $ 3d $ series, interest in the $ 4d $ oxides has 
considerably grown. Prominent examples for such oxides are 
$ {\rm Ca_2RuO_4} $ \cite{mizokawa01}, $ {\rm Sr_2RuO_4} $, which shows 
superconductivity below $ {\rm T_c} \approx 1.5 $\,K \cite{srruo4}, and 
$ {\rm SrRuO_3} $, which is ferromagnetic below 160\,K \cite{srruo3}.

Recently, focus has centered on the new ruthenate $ {\rm La_2RuO_5} $, 
which shows a first-order phase transition near 160\,K \cite{khalifah02}. 
This semiconductor-semiconductor transition is associated with a slight 
increase of the band gap from $ \approx 0.15 $\,eV to about $ 0.21 $\,eV. 
In addition, it is accompanied by strong changes in the magnetic 
properties as well as by a transformation from a monoclinic to a 
triclinic lattice. High-temperature Curie-Weiss behaviour, with
$ \mu_{\rm eff} = 2.53 \mu_B $ and  $ \Theta = - 71 $\,K, is attributed to 
the low-spin  ($ S = 1 $) moments of the $ {\rm Ru^{4+}} $ ions. At the transition, 
the susceptibility drops to a small, nearly temperature-independent 
value with a slight upturn at lowest temperatures assigned to free 
intrinsic spins or extrinsic impurities \cite{khalifah02,malik05}.
Furthermore, from the absence of any field dependence of the magnetic 
susceptibility for fields up to 9\,T a complete quenching of the 
magnetic moments was deduced and attributed to an orbital ordering 
of the Ru ions.

\begin{figure}[tb]
\centering
\includegraphics[width=0.48\textwidth]{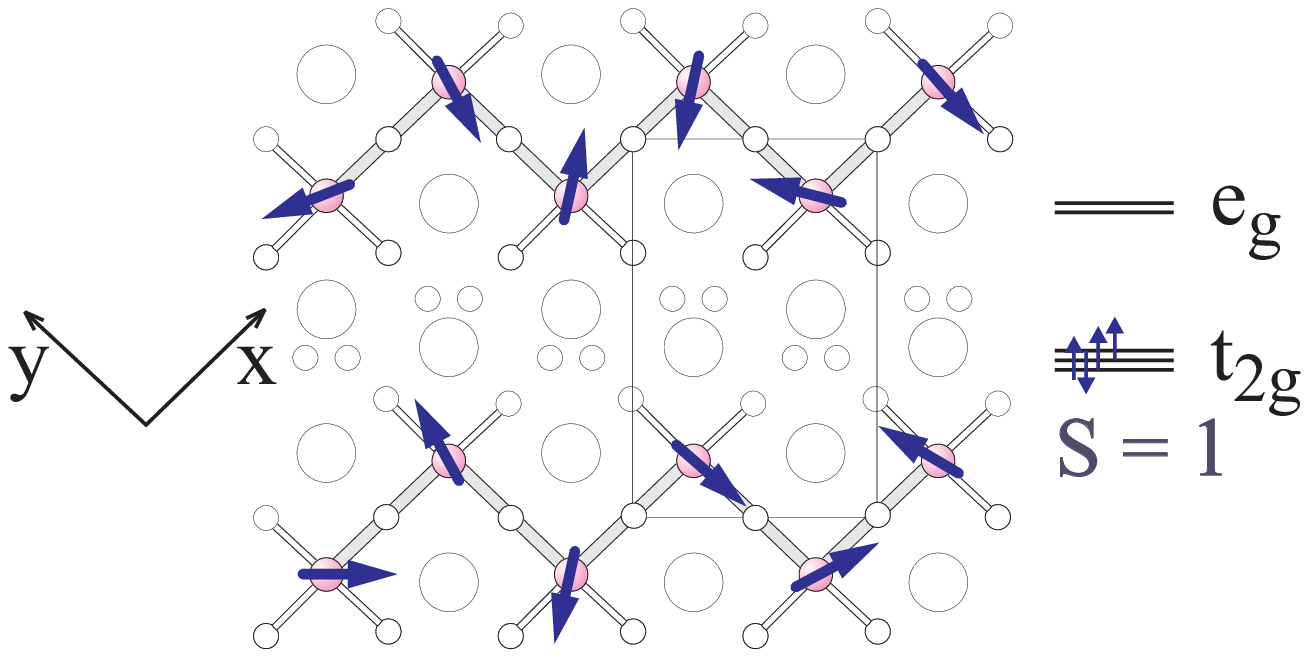}
\caption{Crystal structure of ht-$ {\rm La_2RuO_5} $ viewed
         along the $ c $-direction. $ {\rm La} $, $ {\rm Ru} $, and
         $ {\rm O} $ atoms are displayed as big, medium, and small
         circles, respectively.}
\label{fig:cryst1}
\end{figure}

A schematic representation of the crystal structure is displayed
in Fig.\ \ref{fig:cryst1}. 
Arrows indicate the random orientation of the spin moments in the 
high-temperature (ht-) phase. The structure is built from octahedral 
chains parallel to the $ c $-axis, which 
form a zigzag-like pattern in the $ ab $-plane, thus giving rise to 
double-layer slabs, which are separated by LaO-layers
\cite{khalifah02,boullay03,ebbinghaus05}. The main features of the
crystal structure are preserved in the low-temperature (lt) phase
\cite{khalifah02,ebbinghaus05}. However, according to the neutron
data the magnetic transition is accompanied by pronounced local
structural changes leading to alternating shortenings and
elongations of the Ru--O--Ru bond lengths \cite{ebbinghaus05}.
While the Ru--Ru nearest neighbour distances are quite similar to
each other in the ht-phase, the lt-phase is characterized by short
and long distances alternating both within the plane and along the
$ c $-axis.
The in-plane octahedral Ru--O bonds, which range from 1.94 to
2.06\,\AA \ in the ht-phase, fall into two short and two long
bonds ranging from 1.87 to 1.97\,\AA \ and 1.98 to 2.10\,\AA,
respectively. The bonds parallel to the $ c $-axis are also larger
than 2.0\,\AA. In the rotated coordinate system sketched in Fig.\
\ref{fig:cryst1}, the long and short bonds evolving in the lt-phase 
are along the $ x $- and $ y $-axis, respectively. The structural 
changes lead to two inequivalent Ru sites, which alternate along 
the $ c $-axis as well as the zigzag-like in-plane pattern. 
>From structural considerations, Khalifah {\em et al.}\ concluded 
that in the lt-phase the Ru $ 4d_{yz} $ orbitals are depopulated,
leading to the configuration $ d_{xy}^2 d_{xz}^2 d_{yz}^0 $,
hence, an $ S = 0 $ state. In contrast, in the ht-phase the 
$ t_{2g} $ states would be nearly degenerate and Hund's rule
coupling gives rise to the $ S = 1 $ electron configuration of 
$ d_{xy}^2 d_{xz}^1 d_{yz}^1 $,  $ d_{xy}^1 d_{xz}^2 d_{yz}^1 $, 
or $ d_{xy}^1 d_{xz}^1 d_{yz}^2 $. Thus, the phase transition 
was interpreted as an orbital ordering transition with a complete 
loss of the local magnetic moment. Finally, the occurrence of an
inelastic peak at about 40\,meV as observed in neutron scattering
was assigned to the formation of a spin gap in the lt-phase
\cite{khalifah02}. Underlining the importance of intersite
interactions, Osborn pointed out that the observed
inelastic response should be rather attributed to singlet to
triplet type excitations \cite{osborn04}. This point of view was supported by
Khomskii and Mizokawa \cite{khomskii05a}.

Concentrating on the above mentioned discrepancies in interpreting
the data for the lt-phase, we report on density functional
calculations as based on crystal structure data for both phases
\cite{ebbinghaus05}. As expected, we find strong changes of the
orbital occupations coming with the structural transformation.
However, in contrast to the proposal by Khalifah {\em et al.},
these changes conserve the local $ S = 1 $ moment. From
spin-polarized calculations it is inferred that the suppression of
the susceptibility in the lt-phase results from a
spin-Peierls-like transition coming with the formation of
spin-ladders with antiferromagnetic coupling on the rungs.

%\section{Methodology}

The calculations were performed using the scalar-relativistic
augmented  spherical wave (ASW) method \cite{wkg,revasw}. The
large voids of the open crystal structure were accounted for
by additional augmentation spheres, which were automatically
generated by the sphere geometry optimization (SGO) algorithm
\cite{eyert98b}.
The Brillouin zone sampling was done using an increased number of
up to 1024 and 2048 $ {\bf k} $-points in the irreducible wedge of
the monoclinic and triclinic Brillouin zone, respectively. While
previous calculations (Ref.\ \cite{wop05}) were based
on the local density approximation (LDA), the present work used
additionally the generalized gradient approximation (GGA)
\cite{perdew96a} as well as a new version of the ASW code, which
takes the non-spherical contributions to the charge density inside
the atomic spheres into account \cite{eyerttbp}.

%\section{Results and Discussion}

\begin{figure}[tb]
\centering
\includegraphics[width=0.48\textwidth]{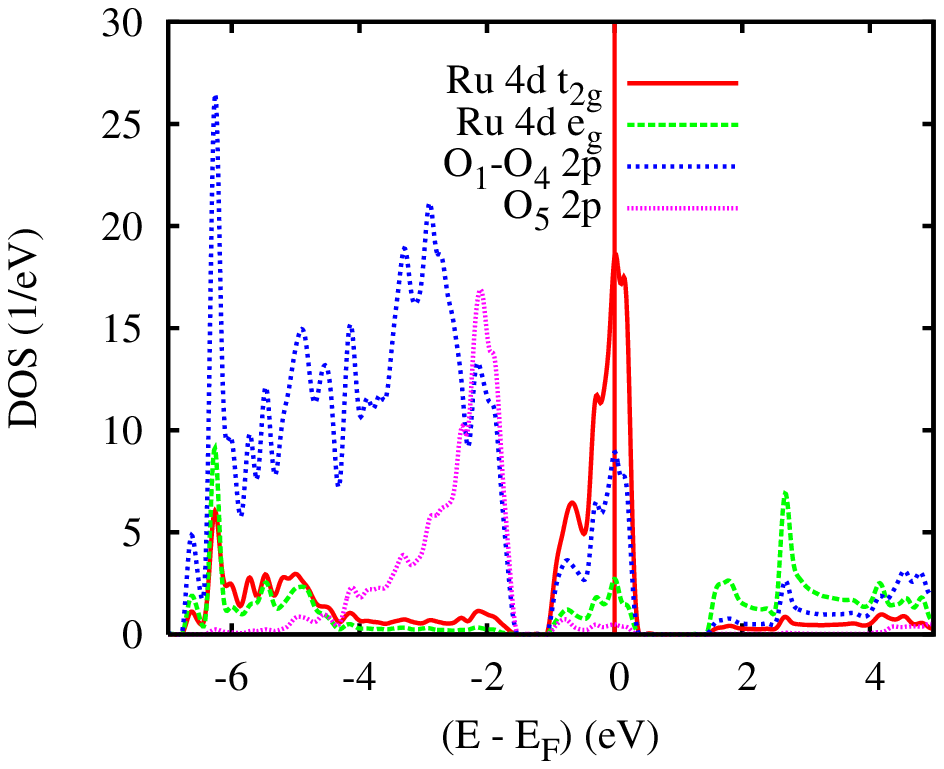}
\includegraphics[width=0.48\textwidth]{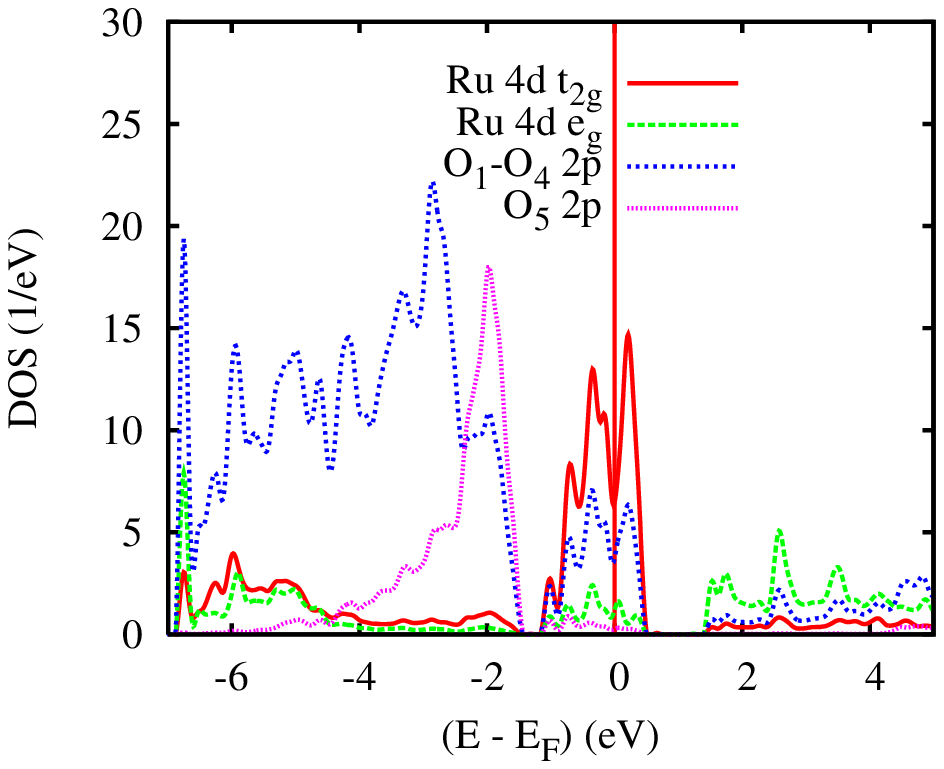}
\caption{Partial DOS as resulting from the ht- (top) and lt- (bottom)
         crystal structure. % Slight broadening is due to the DOS
         % calculation scheme \protect \cite{methfessel89}.
         }
\label{fig:res1}
\end{figure}

Calculated partial densities of states (DOS) are displayed in Fig.\
\ref{fig:res1}
for the high-temperature (ht) and low-temperature (lt) structure.
Contributions from both Ru $ 4d $ and O $ 2p $ states are included.
For the oxygen contributions we have distinguished the atoms O(1)
to O(4), which form the $ {\rm RuO_6} $ octahedra, from those of
atom O(5), which is at the center of the $ {\rm La_4O} $ tetrahedra.
Not shown are the La $ 5d $ and $ 4f $ states, which give rise to
a sharp peak at about $ 5 $\,eV and smaller contributions between
$ -4.5 $ and $ -1.5 $\,eV. All other orbitals play only a negligible
role in the energy interval shown.

In Fig.\ \ref{fig:res1}, the lowest group of bands extending
from $ -6.9 $ to $ -1.5 $\,eV derives mainly from the O $ 2p $
states. In contrast, the Ru $ 4d $ states are found mainly in
the energy interval from $ -1.1 $ to $ 0.5 $\,eV as well as
above $ 1.5 $\,eV. Due to the octahedral coordination of Ru by
oxygen atoms these states are split into $ t_{2g} $ bands near
the Fermi energy and $ e_g $ states well above. Hybridization 
with the O $ 2p $ states is extremely strong and leads to large 
$ p $-admixtures to the DOS near $ {\rm E_F} $, which are of the 
order of $ \approx 40 $\% of the Ru $ 4d $ contributions in that 
interval.

The partial DOS of the O(5) $ 2p $ states deviates considerably
from those of the remaining four oxygen atoms and shows a steady
increase between $ -4.5 $ and $ -1.5 $\,eV as well as a sharp drop
at the upper edge. La $ 5d $ and $ 4f $ admixtures in this
interval of the order of 15\% of the O(5) $ 2p $ contribution are
attributed to strong hybridization coming with the formation of
the $ {\rm La_4O(5)} $ tetrahedra. The large band width of these
rather localized tetrahedral states reflects their considerable
extent in space.

In general, the similarity of the crystal structures of the ht-
and lt-phase is well reflected by the similarity of the partial
DOS. Yet, distinct changes are observed for the Ru $ 4d $ $ t_{2g} $
states. In particular, this group of bands becomes broader and
the strong peak at $ {\rm E_F} $, seen for the ht-structure, turns
into a pronounced dip. Actually, as can be seen on closer
inspection of the band structure, a tiny gap is opened in the
lt-phase.

\begin{figure}[tb]
\centering
\includegraphics[width=0.48\textwidth]{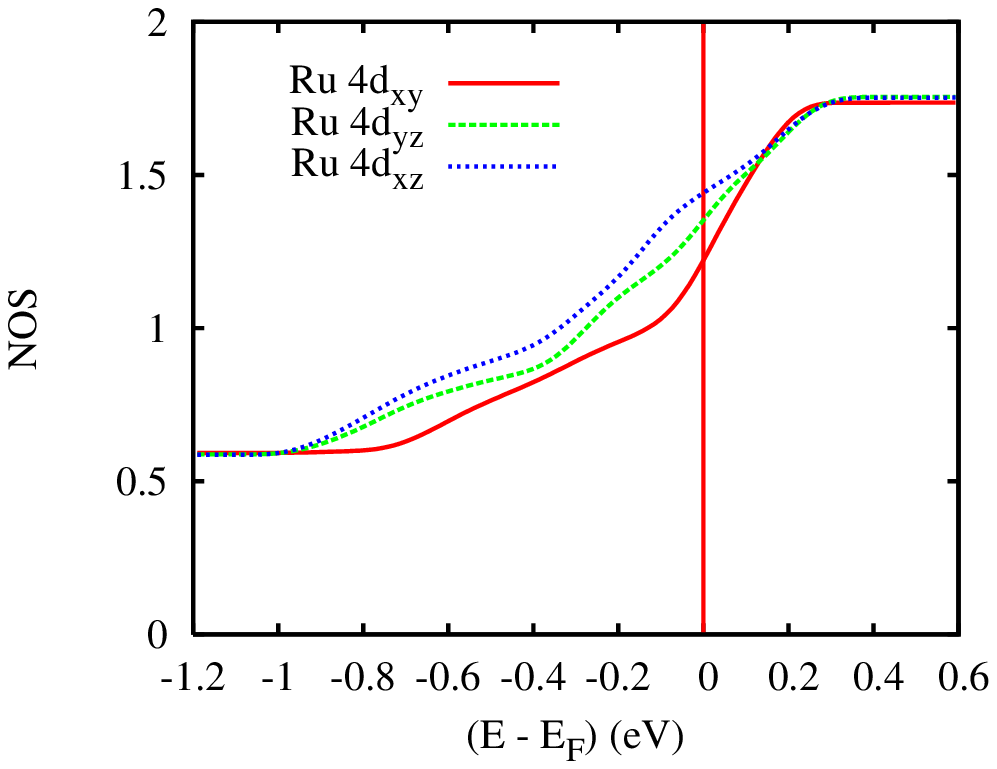}
\includegraphics[width=0.48\textwidth]{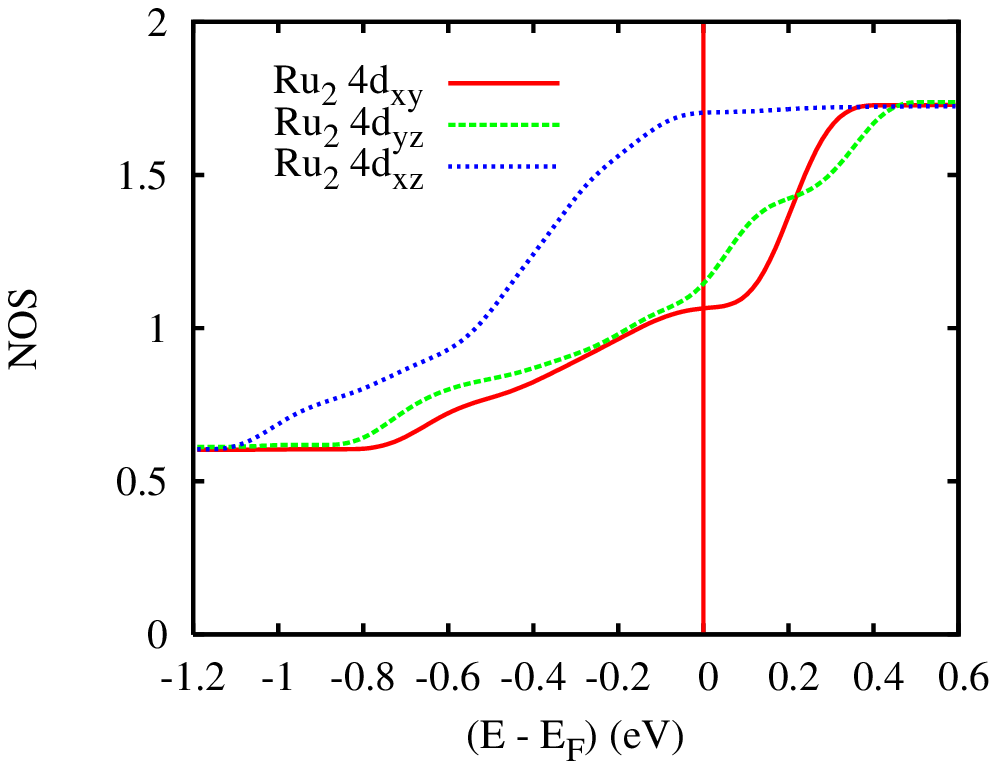}
\caption{Integrated partial Ru $ 4d $ $ t_{2g} $ DOS as resulting from
         the ht- (top) and lt- (bottom) crystal structure. Results
         for the Ru(1) atom of the lt-phase are very similar to those
         for atom Ru(2) as shown here. Orbitals refer to the rotated
         coordinate system depicted in Fig.\ \protect \ref{fig:cryst1}.} 
\label{fig:res2}
\end{figure}

The differences between the electronic properties of the ht- and
lt-phase become much clearer from a detailed analysis of the
near-$ {\rm E_F} $ states. For the notation of the orbitals the
rotated coordinate indicated in Fig.\ \ref{fig:cryst1} is used. In
this system, the partial densities of states of all three Ru $
t_{2g} $ bands display the expected rather similar behaviour for
the ht-phase. As a consequence, integration these partial DOS
leads to almost identical orbital occupations as shown in the
upper panel of Fig.~\ref{fig:res2}.
Note that the $ t_{2g} $ occupations do not vary from 0 to 2 in
the energy interval shown as would be expected from pure $ d $
states. Due to the strong $ p $-admixture to these bands the
energy variation of the orbital occupations appears to be somewhat
reduced and eventually has to be translated to the ideal picture
of pure $ d $-states. Obviously, filling the nearly degenerate $
t_{2g} $ bands according to Hund's rules leads to one of the
electronic configurations $ d_{xy}^2 d_{xz}^1 d_{yz}^1 $, $
d_{xy}^1 d_{xz}^2 d_{yz}^1 $, or  $ d_{xy}^1 d_{xz}^1 d_{yz}^2 $
as proposed in Ref.\ \cite{khalifah02}.

A noticeably different situation is obtained for the lt-structure.
Integrated partial densities of states for the atom Ru(2) are
shown in the lower panel of Fig.\ \ref{fig:res2}. The
corresponding curves for atom Ru(1) are very similar and thus not
displayed. Strong similarities between the $ d_{xy} $ and $ d_{yz}
$ partial DOS are observed.
In contrast, the $ d_{xz} $ state deviate substantially. In
particular, occupation of this orbital is much larger as compared
to that of the other states as is expected from the elongation of
the Ru--O-bonds along the local $ x $-axis. Transferring these
findings to the ideal picture of pure $ d $ states leads to the
electronic configuration $ d_{xy}^1 d_{xz}^2 d_{yz}^1 $. This is
in strong contradiction to the $ d_{xy}^2 d_{xz}^2 d_{yz}^0 $
state proposed by Khalifah {\em et al.}, who claim the full
occupancy of the $ d_{xy} $ orbital at the expense of the $ d_{yz}
$ state. Yet, the similar occupations of the $ d_{xy} $ and $
d_{yz} $ orbitals obtained from the present calculations are more
in line with bond-length considerations using the Ru--O distances
discussed above. Our results have important consequences for the
magnetic moments. While Khalifah {\em et al.}\ propose the
complete quenching of the local moment, the present calculations
clearly reveal the conservation of the $ S = 1 $-spin moment
through the phase transition. However, since the orbital
degeneracy has been lifted by the structural distortion coming
with the triclinic phase, this moment is carried exclusively by
the $ d_{xy} $ and $ d_{yz} $ orbitals.

The identification of an $ S = 1 $-state for the triclinic structure 
motivated additional spin-polarized calculations for the lt-phase. 
Long-range ferromagnetic order can be ruled out from the low-temperature 
susceptibility data. In addition, the fact that no extra reflections 
were detected in the neutron diffraction data for the lt-phase allows only
for antiferromagnetic order with opposite moments at the inequivalent
Ru sites.
This situation suggests a singlet ground state which, with the
discussed structural and orbital transitions, may be realized as a
spin-Peierls-like state. We have simulated this state by starting
from opposite moments at the Ru(1) and Ru(2) sites. 
Indeed, our calculations resulted in a self-consistent solution, which comes
with an energy lowering of $ 5 $\,meV per Ru pair as compared to
the spin-degenerate case. In addition, an optical band gap of $
0.17 $\,eV is obtained in LDA. GGA calculations result in $ 0.20
$\, eV, which is very close to the experimental value.

The calculated magnetic moments arise to equal parts from the 
$ d_{xy} $ and $ d_{yz} $ orbitals thus confirming the expectation
from the previous spin-degenerate calculations. In total, local
moments of $ 0.77 \mu_B $ and $ -0.73 \mu_B $ are obtained at the
Ru(1) and Ru(2) site, respectively. Together with small
contributions from the nearest neighbour oxygen atoms a magnetic
moment of $ \pm 0.85 \mu_B $ per octahedron results, which
increases to $ \pm 1.06 \mu_B $ per octahedron on going from LDA
to GGA. Interestingly, despite the absence of any symmetry
constraint, both types of calculations lead to a
compensation of the magnetic moments at neighbouring octahedra
and, hence, to an exactly vanishing magnetic moment per unit cell.
This finding is in excellent agreement with the
suppression of the magnetic susceptibility below the phase
transition.

\begin{figure}[tb]
\centering
\includegraphics[width=0.48\textwidth]{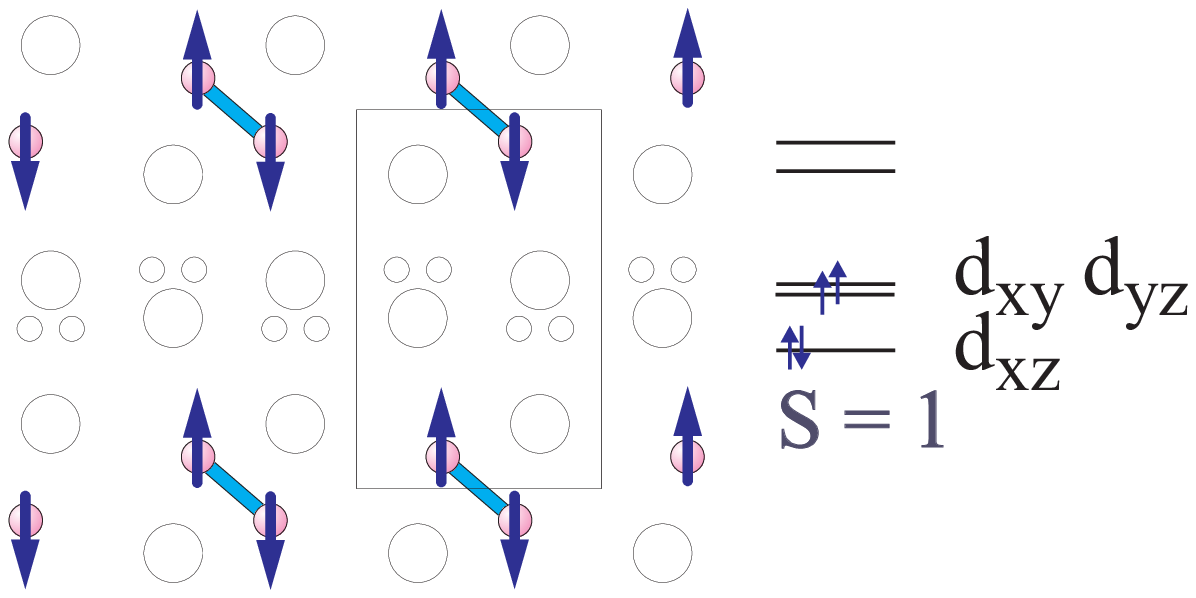}
\caption{Crystal and spin structure of lt-$ {\rm La_2RuO_5} $ viewed
         along the $ c $-direction. $ {\rm La} $, $ {\rm Ru} $, and
         $ {\rm O} $ atoms are displayed as big, medium, and small
         circles, respectively.}
\label{fig:cryst2}
\end{figure}

Finally, we performed spin-polarized calculations also for the
ht-phase, assuming the same type of ordering as for the lt-structure.
As a result, a metallic solution with magnetic moments of
$ \pm 0.76 \mu_B $ per Ru atom was obtained, however, with a total
energy much higher than that of the corresponding spin-degenerate
solution. Taking into account both the GGA and the non-spherical
contributions we arrived at moments of $ \pm 1.11 \mu_B $ per Ru
atom and an optical band gap of $ 0.02 $\, eV. The energetical
instability of this solution agrees with the fact that no long-range
magnetic order is observed for the ht-phase. The phase transition to
the lt-phase may thus be regarded as a spin-Peierls-like transition.
Since the octahedra form chains parallel to the $ c $-axis, the
structural transformation leads to the formation of spin ladders
with antiferromagnetic coupling along the rungs as indicated in Fig.\
\ref{fig:cryst2}.
%

%\section{Conclusion}

To conclude, electronic structure calculations for the ht- and
lt-structure of $ {\rm La_2RuO_5} $ reveal strong orbital ordering
for the latter. While Hund's rule coupling within the degenerate 
$ t_{2g} $ manifold leads to a $ d_{xy}^2 d_{xz}^1 d_{yz}^1 $, 
$ d_{xy}^1 d_{xz}^2 d_{yz}^1 $, or  $ d_{xy}^1 d_{xz}^1 d_{yz}^2 $
state in the ht-phase, the structural changes associated with the
triclinic structure cause substantial orbital ordering and drive
the system into a $ d_{xy}^1 d_{xz}^2 d_{yz}^1 $ configuration.
While well reflecting the Ru--O bond lengths the latter preserves
the local $ S = 1 $-moment and thus is in contradiction with
previous interpretations of the lt-phase.
Spin-polarized calculations for the low-temperature phase reveal
compensation of the local moments due to their
antiparallel alignment along the short Ru--Ru bonds.
As the formation of short and long in-plane bonds in the triclinic
phase leads to an effective pairing of chains, a spin-ladder system
is generated.

We are grateful to I.\ Leonov, J.\ Rodriguez-Carvajal,
and C.\ Schuster for fruitful discussions. This work was supported
by the Deutsche Forschungsgemeinschaft (DFG) through
Sonderforschungsbereich SFB 484 and by the BMBF 13N6918 (T.K.).

\end{document}